\documentclass[12pt,preprint]{aastex}
\begin{document}
%\received{}
%\accepted{}
%\revised{}
%\lefthead{}
%\righthead{}
\shorttitle{Comparative Abundances Analyses} \shortauthors{Williams et
al.}  
\title{Comparative Absorption and Emission Abundance Analyses of
Nebulae: Ion Emission Densities for IC 418 \footnote{Based on
observations made with the NASA/ESA Hubble Space Telescope, obtained
from the STScI, which is operated by AURA, Inc. under NASA contract
NAS 5-26555, and on observations made at CTIO/NOAO, which is operated
by AURA, Inc. under cooperative agreement with the NSF.  }}

\author{Robert Williams} 
\affil{Space Telescope Science Institute,
3700 San Martin Drive, Baltimore, MD 21218} \email{wms@stsci.edu}

\author{Edward B. Jenkins}
\affil{Princeton University Observatory, Princeton, NJ 08544}
\email{ebj@astro.princeton.edu}

\and

\author{Jack A. Baldwin and Brian Sharpee} \affil{Department of
Physics and Astronomy, Michigan State University, East Lansing, MI
48824} \email{baldwin@pa.msu.edu, sharpee@pa.msu.edu}

\begin{abstract}
Recent analyses of nebular spectra have resulted in discrepant
abundances from CNO forbidden and recombination lines.  We consider
independent methods of determining ion abundances for emission
nebulae, comparing ion emission measures with column densities derived
from resonance absorption lines viewed against the central star
continuum.  Separate analyses of the nebular emission lines and the
stellar UV absorption lines yield independent abundances for ions, and
their ratio can be expressed in terms of a parameter
$<\!\!n_e\!\!>_{em}$, the ``emission density'' for each ion.  Adequate
data for this technique are still scarce, but separate analyses of
spectra of the planetary nebula and central star of IC 418 do show
discrepant abundances for several ions, especially Fe II.  The
discrepancies are probably due to the presence of absorbing gas which
does not emit and/or to uncertain atomic data and excitation
processes, and they demonstrate the importance of applying the
technique of combining emission- and absorption-line data in deriving
abundances for nebulae.
\end{abstract}


\section{Introduction}

Much of our knowledge of physical conditions and abundances in diffuse
emission regions such as \ion{H}{2} regions, planetary nebulae (PNe),
and supernova remnants derives from an analysis of the emission-line
spectra.  The results from studying these objects, particularly
\ion{H}{2} regions, are then key ingredients in the study of abundance
gradients on galactic scales and as a function of lookback time. The
methods of analysis have been well developed over the past half
century, and have been applied to virtually every type of
emission-line object \citep{O89,DS02}.  The strongest lines of H and
He are due to electron recapture to higher levels whereas
collisionally excited forbidden lines from lower levels tend to
dominate the heavy element spectrum, and these lines form the
cornerstone of most spectral analyses of nebulae.

It became evident from some of the early IUE UV spectra of PNe that
the relative intensities of the stronger recombination and
collsionally excited lines originating from the same ion were in
disagreement with theory and that some additional process might be the
cause \citep{HLS81}.  Similar types of discrepancies have arisen in
the past and various processes have been invoked to explain them,
e.g., fluctuations in electron temperature \citep{P67} and density
\citep{V94}, fluorescence scattering of continuum radiation
\citep{S68}, and selective excitation mechanisms involving charge
transfer \citep{W73} and dielectronic recombination \citep{S81}, with
some success achieved for selected lines.

Recently, however, relative intensities of a large sample of CNO
forbidden and recombination lines have been determined from high
signal-to-noise spectroscopy of PNe, and the problem of discrepant
forbidden and recombination line intensities within the same ion has
persisted (Liu et al. 1995, 2000; Garnett \& Dinerstein 2001a) .  Some
objects exhibit the problem, while others do not, and its
manifestation is the variation in the intensities of an ion's
recombination lines as a whole relative to the forbidden lines of that
ion among different objects.  No convincing explanation has emerged
that accounts for this phenomenon (Garnett \& Dinerstein 2001b; Liu
2002a, 2002b).  Analyses of optical and IR spectra of various PNe have
now extended the problem generally to CNO ions and to \ion{H}{2}
regions \citep{P93,E99,T02}, and the discrepancies between the
predicted and observed intensities of forbidden lines relative to the
recombination lines from the same ion in selected nebulae can exceed
an order of magnitude.  The consequence of this situation is that
abundances derived from the emission lines may be incorrect, and it is
not clear which lines produce reliable results.

There have been additional signs of problems related to the excitation
of forbidden lines in elements heavier than CNONe.  It was already
established some years ago that the observed fluxes of [\ion{Fe}{2}]
and [\ion{Ca}{2}] emission in nebulae are frequently orders of
magnitude weaker than predicted from model calculations
\citep{S78,K95}, and this has been attributed to selective gas
depletion of these elements by condensation into dust grains.  In the
other direction, strengths of certain lines observed in emission
objects, such as [\ion{Fe}{10}] $\lambda$6375 and [\ion{Ni}{2}]
$\lambda$7378, often greatly exceed their expected, calculated
intensity relative to other ions \citep{F84,H88}.  This phenomenon has
been attributed to the selective excitation of the lines from
scattering of stellar continuum radiation by a strong resonance
transition that populates the upper level of the forbidden line
\citep{K89,LU95}.

The discrepancies that persist between predicted and observed line
strengths demonstrates that assumed excitation processes may be
incorrect and, in turn, some abundances derived from emission-line
analyses may not be valid.  Element abundances and physical conditions
that are derived from emission lines are fundamental to our
understanding of diffuse gas, and an independent method of determining
abundances is essential to elucidate the present situation.  An
alternative method of deriving ion abundances does exist and is
frequently applied to non-emitting regions of the ISM and IGM:
analysis of the absorption lines produced by the gas against a more
distant continuum source.  For ions whose absorption lines are not
strongly saturated column densities can be determined relatively free
of uncertain parameters, and this technique can provide an independent
check of emission line results.

Absorption analyses of emission nebulae have been performed for a
relatively small number of objects, and studies for which the results
have been compared with an emission analysis of the same ions in the
same object are virtually non-existent.  \citet{J76} and \citet{S80}
recognized that the Vela and SN 1006 supernovae remnants are seen
against a background of more distant stars and they obtained high
resolution spectra of some of the stars in the UV and optical,
identifying various ion species produced by the gas of the SNR's.  The
first abundance determinations of nebular gas in PNe from absorption
lines were attempted by \citet{PMP84} and \citet{PPM86} from high
resolution UV specta of the central stars of NGC 6543 and
BD+30$^o$3639 obtained with IUE.  They derived column densities for
certain ions and compared them to those inferred from relative
emission line intensities, finding general agreement between the
emission and absorption strengths.  \citet{D95} independently surveyed
neutral gas outside the H$^+$ zone in PNe shells by observing the
\ion{Na}{1} D absorption lines in the continua of their central stars,
and made positive detections in more than 30\% of their sample from
which they deduced column densities.  \citet{W02} determined column
densities for the eta Carinae nebulosity by observing the UV
absorption lines produced by the ejecta in the spectra of more distant
hot stars lying behind it in the galactic plane.  In none of the above
studies were absolute emission intensities of the nebular lines used
to determine emission measures for the same ions as those for which
column densities had been determined from the absorption lines.

We present here a technique that can be used to determine ion
abundances in nebulae, and which makes independent use of both (a) the
resonance absorption lines produced in the central star continuum by
the shell of the nebula, and (b) the emission lines from the nebula
along an adjacent line of sight.  Unfortunately, good signal/noise
data for emission and absorption lines from the same ion in the same
object currently exist for only one nebula: the PN IC 418.  We
determine comparative abundance ratios using the two methods of
analysis for several ions in this nebula which demonstrate the value
of combining the two methods.  The results reveal inconsistencies
between the two methods that cast some doubt on the usual assumptions
governing the excitation of certain forbidden lines.

\section{The Emission Density}

The observed extinction-corrected flux of an optically thin forbidden
line can be written
\begin{equation}
\label{eqn1}
F_{\lambda} = {\theta_o}^2 \int n_e \,n_i \, q(T_e) \, d\ell = {\theta_o}^2
\zeta \int^{0}_{-R} n_e \, n_i \, q(T_e) \, d\ell\,,
\end{equation}
where ${\theta_o}^2$ is the angular area of the gas being observed,
$q(T_e)$ is the line excitation coefficient appropriate for the
transition, $R$ is the nebula radius, and $n_e$ and $n_i$ are the
electron and ion densities.  The first integral in equation~\ref{eqn1}
is taken over the entire emission line of sight through the nebula and
is the emission measure, $EM_i$, of the ion times the coefficient
$q(T_e)$.  The parameter $\zeta$ is a geometrical factor that converts
the line integral through half the nebula to the star to the
appropriate integral for the observations, which sample a longer path
through the entire nebula offset from the central star.  The factor
$\zeta$ thus normalizes the emission and absorption line strengths to
the same path length for comparison purposes, and its value, normally
of order unity, depends on the geometry of the nebula and the
placement of the spectrograph slit, as will be discussed further in
\S~\ref{emission}.

For most forbidden lines $q(T_e)$ is the collisional excitation
coefficient corrected for any collisional de-excitation, which
consists of contributions from multiple levels but is almost always
dominated by direct impact excitation of the upper level of the line
from the ground state.  Under some circumstances $q(T_e)$ consists of
additional terms when other processes such as dielectronic
recombination or resonance fluorescence of stellar continuum radiation
contribute to the line excitation, and in this case the additional
terms can cause $q(T_e)$ to have a density dependence.

The column density of an ion, $N_i$, and the intensities of its
emission lines differ in their dependence upon gas density and
temperature.  It is straightforward to determine both the column
density and the emission measure of an ion from observations of the
absorption line equivalent widths and the absolute fluxes of the
emission lines.  A useful way to relate the absorption and emission
line strengths from an ion is to take the ratio of the ion's emission
measure to its column density.  This ratio defines a density-weighted
mean electron density for that ion, $<\!\!n_e\!\!>_{em}$ , which we
designate as the emission density of the ion,
\begin{equation}
\label{eqn2}
<\!\!n_e\!\!>_{em} \equiv EM_i/(\zeta \, N_i) = F_{\lambda} \left[ \zeta \,\, {\theta_o}^2 \,
q(T_e) \int_{-R}^{0} n_i \, d\ell \right]^{-1}\,,
\end{equation}
and where we have removed the emission coefficient from the integral
by having the line emission from individual ions be represented by
single electron temperatures.

The physical interpretation of the emission density is that it is the
mean electron density required to produce the observed emission line
fluxes of an ion \textit{given the column density of that ion} as
determined from its absorption lines.  Its value can be determined for
an ion from the emission and absorption lines, for an assumed electron
temperature, although inhomogeneities in the gas can produce
departures of the emission density from the electron density if the
absorption and emission lines of sight are dissimilar.  Generally,
however, the $<\!\!n_e\!\!>_{em}$ derived from the emission and
absorption lines should take on values close to that of the local
electron density of the emitting gas.

In making observations of absorption features against the continuum of
the ionizing star, one samples a column of gas having a cross section
equal to that of the star, normally $<10^{12}$ cm across and
frequently orders of magnitude smaller.  However, to obtain good
quality spectra of diffuse emission regions usually requires
observations which sample more than 1\arcsec\ along the slit, which
for an object at 1 kpc distance corresponds to a size $>10^{16}$ cm.
Thus, emission lines of sight sample much larger cross sections than
absorption observations.  An asymmetrical distribution of gas, such as
jets or unusual clouds that intersect one of the emission or
absorption lines of sight but not the other, can clearly lead to
values of the emission density for an ion that differs from the
electron density because the emission and absorption sample gas having
different characteristics.  However, the values of
$<\!\!n_e\!\!>_{em}$ for two ions whose elements have very similar
ionization potentials should be the same, irrespective of
inhomogeneities or the geometry of the object, because such ions
should be co-spatial and share the same physical conditions.  A
comparison of the emission densities for two similar ions is a
valuable check on the assumed excitation processes for their emission
lines.

\section{Observations of the Planetary Nebula IC 418}

Data that would allow abundance determinations from both absorption
and emission lines exist for very few objects.  With the exception of
\ion{Ca}{2}, which has forbidden and resonance lines in the visible,
no other abundant ion has both resonance lines and observable emission
lines in the visible spectral region.  Most absorption lines produced
by diffuse gas originate from the ground state and occur in the UV, so
space observations are necessary unless the object has large redshift.
The local ISM produces strong absorption troughs from many of the same
transitions that are formed in Galactic emission regions, hence the
detection of discrete absorption components from any object requires
that it have a different radial velocity than that of the ISM along
the line of sight.  The best candidate objects to observe are PNe
because many of them have relatively bright central stars in the UV,
relatively high surface brightness emission shells, and radial
velocities that differ from that of the intervening disk ISM.  The
compact PN IC 418 satisfies all these criteria and its central star
has already been observed at high spectral resolution with {\sc
HST}/{\sc STIS}, with the data publicly available from the {\sc HST}
archive\footnote{The data are available via
http://archive.stsci.edu/cgi-bin/nph-hst as Datasets 052902010, -030,
and -040 (PI: H. Dinerstein).}.  We had recently embarked upon a
program to obtain high signal/noise optical spectra of selected PNe
which included IC 418, thus relevant data to determine ion abundances
separately from emission and absorption already exist for this one
object.

\subsection{Ultraviolet Absorption Features}

The central star of IC 418 was observed in the UV in 1999 March with
{\sc HST}/{\sc STIS} over the wavelength region 1150-1687\AA\ at a
resolution of 8 km s$^{-1}$.  We obtained the on-the-fly calibrated
data on 2002 January 4 from the {\sc HST} archive running CALSTIS
version 2.12a, which includes a two-dimensional correction for
scattered light \citep{L98}.  The spectrum consists of a stellar
continuum interspersed with discrete absorption features, the great
majority of which are produced by gas in the intervening ISM.  At a
distance of roughly 1 kpc IC 418 has galactic coordinates of (215,
-24) deg and a heliocentric radial velocity of +60 km s$^{-1}$, so
absorption produced by the nebular shell is adequately
velocity-shifted away from all but the strongest and broadest of the
corresponding ISM absorption features.

\begin{figure}
\figurenum{1} \plotone{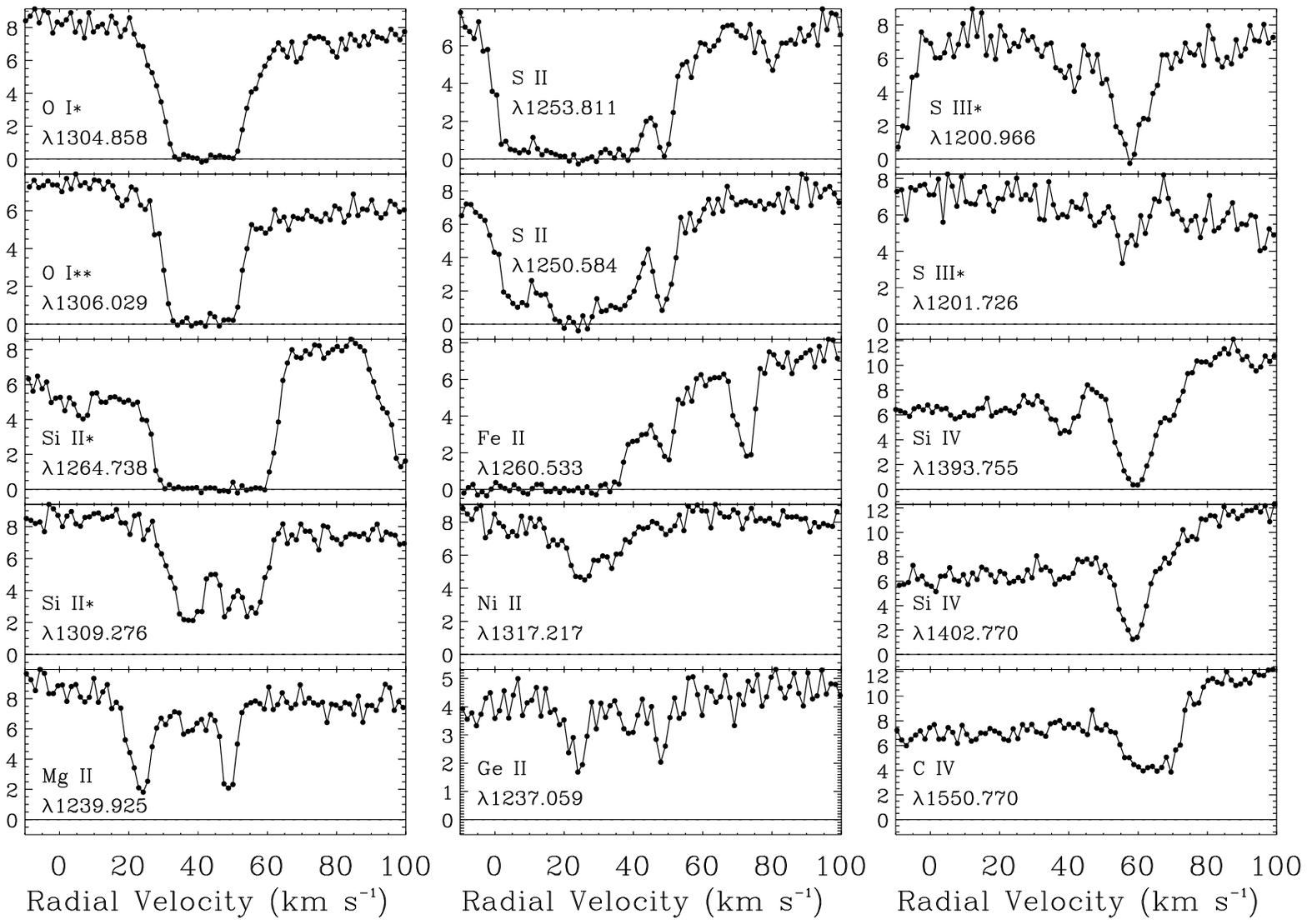} \figcaption[fg1.eps]{ Samples of
absorption features in the {\sc STIS} spectrum of IC 418.  The two
left-hand columns of plots represent species in low ionization stages,
while the right-hand column shows more highly ionized species.  Pairs
of lines from a given ion are arranged such that the stronger
transition appears above the weaker one.  The feature whose left-hand
edge can be seen to the right of the \ion{Si}{2}* $\lambda$1264.738
absorption is a weaker transition of \ion{Si}{2}* at
$\lambda$1265.002.  The y-axis numbers represent flux in units of
$10^{-12}$ erg cm$^{-2}$ s$^{-1}$ \AA$^{-1}$.  \label{fig1}}
\end{figure}

We have identified the nebular absorption lines that appear in the
spectrum, specifically noting unblended lines at the PN shell velocity
that belong to ions which also have emission lines observed in the
visible.  A number of relevant absorption features are either badly
saturated or blended with stronger ISM components and cannot provide
column densities.  For this reason the sample of ions for our combined
analysis is limited to just the five species \ion{O}{1}, \ion{S}{2},
\ion{S}{3}, \ion{Fe}{2}, and \ion{Ni}{2}, although nebular absorption
from other ions is present.  Moreover, there are several velocity
components near the velocity of the shell, as best shown by the
profile of \ion{Si}{2}* $\lambda$1309.276\AA\ (cf. Fig.~\ref{fig1}).
Note that a line denoted by one or two asterisks refers to a
transition originating from the first or second excited fine-structure
level, respectively, within the ground state.

Absorption features whose velocities are in the vicinity of v=+45 km
s$^{-1}$ are assumed to be associated with the foreground part of the
expanding (at $\sim 15$ km s$^{-1}$) shell of IC 418.  The list of
detected absorption lines having a nebular origin is given in
Table~\ref{tbl1} together with our measurements of the equivalent
widths of these lines, the relevant atomic parameters of the
transitions, and the resultant column densities when determinable.
Plots of most of the absorption features are shown in Fig.~\ref{fig1}.
The entries for the elements in the table are ordered by their rank in
the periodic table and according to line strength (strongest line
first), while the panels in the Figure are segregated by ionization.
In the Figure, there is a clear progression toward somewhat higher
velocities for the more highly ionized species, with \ion{C}{4}
reaching a velocity of about v $\approx$ +60 km s$^{-1}$.  This is
expected in a differentially expanding shell in which the lower
ionization gas has the larger expansion velocity, and our emission
line data are entirely consistent with this.  They were taken with a
slit set off-center on the nebula, and inspection of the
two-dimensional long-slit spectra clearly shows that the low
ionization lines come from the outer regions with larger expansion
velocity, while the higher ionization lines originate closer to the
ionizing star and have a lower expansion velocity relative to the
star.

To measure equivalent widths we used continuum levels defined by least
squares fits of Legendre polynomials to intensities on either side of
each line, following the methods described by \citet{S92}.  The
principal foreground interstellar absorption features are found at
v=+25 km s$^{-1}$, and for \ion{S}{2} and \ion{Fe}{2} they overlap the
components associated with IC 418.  The continuum levels for these
lines were defined by the intensities along the apparent edges of the
foreground lines, and thus are not as certain.  The 1-$\sigma$ errors
listed in Table~\ref{tbl1} reflect the combined uncertainties arising
from two effects: one is the uncertainty of the defined continuum
level, and the other is the error produced by random noise for
intensities within the wavelength interval of the line.  These effects
should be independent of each other, hence their magnitudes can be
added in quadrature to arrive at a final error estimate.

The circumstances under which the column densities have been
calculated differ considerably for the different species, reflecting
the large disparities in line strengths.  We summarize the four
approaches we used for deriving $\log N_i$, together with the
applicable ions in each case:

\begin{enumerate}
\item The line is undetected or marginally detected: the listed upper
limit equals the formally derived value plus a 2$\sigma$ error
(\ion{Fe}{2} $\lambda$1611, \ion{Fe}{2}*, \ion{Ni}{2}).

\item A single line of moderate strength is measured: the listed value
assumes no saturation, so the true amount could be somewhat greater if
there are unresolved saturated components present (\ion{C}{4},
\ion{Fe}{2} $\lambda$1261, \ion{Ge}{2}).

\item Two lines are measured and they are probably saturated: the
listed limits correspond to a standard doublet-ratio calculation
assuming the lower limit for the column density is given by the weak
line's W$_{\lambda} -\!\!1\sigma$ and the strong line's W$_{\lambda}
+\!\!1\sigma$, while the upper limit is given by the weak line's
W$_{\lambda} +\!\!1\sigma$ and the strong line's W$_{\lambda}
-\!\!1\sigma$ (\ion{Mg}{2}, \ion{Si}{4}, \ion{S}{2}, \ion{S}{3}**).

\item The line is flat-bottomed and heavily saturated.  A lower limit
is derived assuming $\tau_o=4$ (\ion{O}{1}*,\ion{O}{1}**).
\end{enumerate}

Although two lines of \ion{Fe}{2} lie within our observed spectral
range, we could only show that the stronger line is not saturated
since the other line is weaker by a factor of 14, and is not detected.
The column density range for \ion{S}{3}* was calculated assuming the
velocity dispersion was identical to each of the two solutions we
found by performing the doublet ratio analysis of \ion{S}{3}**.  The
upper limit of infinity arises from the fact that the upper limit
calculation for \ion{S}{3}** gives a value b=1.1 km s$^{-1}$ that
makes the \ion{S}{3}* line too saturated to give a reliable column
density limit.  We emphasize that for ions having multiple levels
within the ground term the derived column densities in
Table~\ref{tbl1} apply only to those ions occupying the specific lower
level of the transition listed.  The total column density of the ion
must account for ions occupying other levels within the ground state.

\subsection{Nebular Emission Lines} \label{emission}

\begin{figure}
\figurenum{2} \epsscale{0.6} \plotone{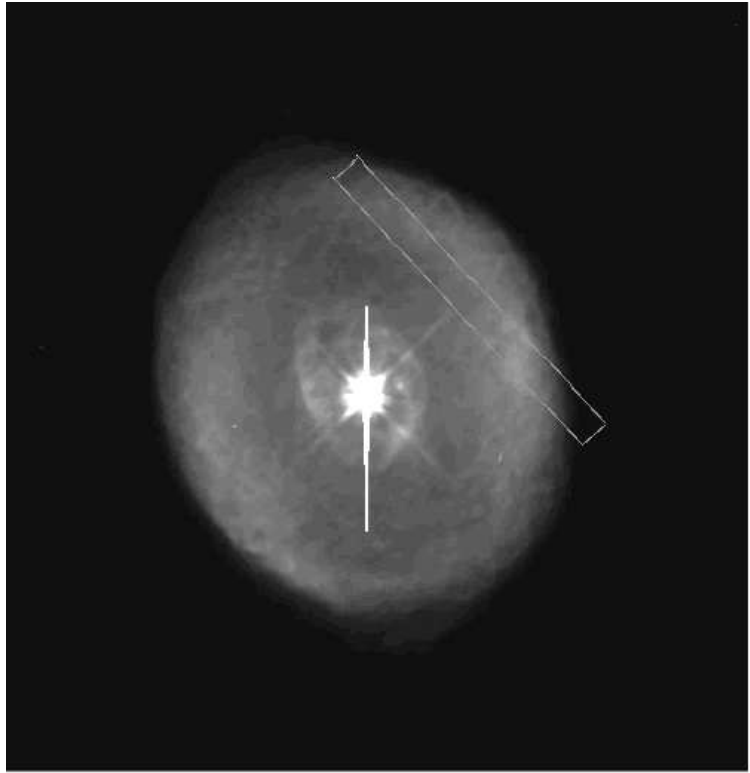} \figcaption[fg2.ps]{ An
{\sc HST} image of IC 418 showing the position and size of the slit
used for the CTIO echelle observations. \label{fig2}}
\end{figure}

As part of a detailed study of faint emission lines in nebulae, we
obtained high resolution spectra of a section of the nebular shell of
IC 418 in 2001 December using the CTIO 4m echelle spectrograph with
the long camera in order to measure relative emission line
intensities.  The observations were made at a spectral resolution of 9
km s$^{-1}$ through a 1\arcsec\ $\times$ 11\arcsec\ slit oriented N-S,
i.e., roughly parallel to the major axis of the nebula, and offset
4.3\arcsec\ to the west of the central star, as shown in
Fig.~\ref{fig2}.  It should be emphasized that this line of sight is
offset from that for which the absorption lines were observed, and
therefore a comparison of emission and absorption results should bear
in mind differences that may be due to inhomogeneities.  A
sufficiently high signal to noise was obtained over five nights of
observing to record fluxes for lines between 3500-9850\AA\ having
intensities down to 10$^{-5}$ that of H$\beta$.  Standard
spectroscopic reduction techniques were applied to produce extracted
flux calibrated spectra, with special care taken to subtract out
echelle ghost and airglow features and system noise in order to
achieve the faintest possible detection limits.  Night sky lines were
removed by using a combination of criteria that enabled such features
to be characterized, including their presence on the portion of the
spectral image that extended beyond the shell of IC 418, their narrow
profiles, and their known wavelengths from the airglow atlas of
\citet{O96}.  A complete description of the data processing and the
resultant emission line fluxes is presented by \citet{S02}.  For the
present study the emission lines from those ions for which we were
also able to identify resonance absorption in the stellar continuum
are listed in Table~\ref{tbl2}, together with the specific
transitions, the extinction-corrected emission fluxes, and atomic
parameters relevant to electron impact excitation of the lines.  The
fluxes refer to the region of the PN shell, which has roughly constant
surface brightness, that was sampled by our 1\arcsec \ $\times$
11\arcsec\ slit.  Although we observed many forbidden and
recombination lines the only published atomic data we could find for
transitions from ions with reliable column densities were for the
brighter forbidden lines listed, and these should be predominantly
collisionally excited.

The emission measures of the ions have been calculated in
Table~\ref{tbl2} from eq.~\ref{eqn1} based upon the
reddening-corrected absolute fluxes of the lines, using the reddening
curve of \citet{S79} with an absolute extinction at H$\beta$ of
$10^{0.35}$ (nebular extinction parameter $c_{\beta}=0.35$, as
determined from the ratios of Balmer and Paschen lines originating
from the same upper level), and taking collisional excitation to be
the primary mechanism of excitation for an assumed 3-level atom.  The
latter assumption should provide reasonably accurate excitation rates,
although the inclusion of additional levels will serve to increase the
line excitation coefficient, especially for a complicated ion such as
\ion{Fe}{2}.  The $EM_i$ were calculated using the atomic parameters
listed in Table~\ref{tbl2} (cf.\ Pradhan \& Zhang 1993; Bautista \&
Pradhan 1996) and electron temperatures of $T_e=(5, 8, \textrm{\&}
10)\times 10^3$K, respectively, for the neutral, singly, and doubly
ionized ions, and the results are given in column 5.  The temperatures
are taken from the values determined for IC 418 by \citet{HA94} from
their spectroscopic study and $n_e/T_e$ solutions for the forbidden
line intensities.  The resultant emission density
$<\!\!n_e\!\!>_{em}$, for each ion has been computed from the ratio of
the emission measure to the column density for the ion, and these are
listed in the final column of Table~\ref{tbl2}.

The ion column densities used to calculate the emission densities have
been determined from the data in Table~\ref{tbl1}, with a correction
for the population of all fine structure levels within the ground
state.  The corrections are as follows for the different ions:

\begin{enumerate}
\item \ion{O}{1} was assumed to have a Boltzmann population among the
ground levels because of the large column densities determined for the
first and second excited levels, and their critical densities of
$n_c$=10$^4$ cm$^{-3}$.

\item \ion{S}{2} has only the one ground level.

\item \ion{S}{3} was also assumed to have a Boltzmann distribution
because of the large column densities observed for the first and
second excited levels, and their critical densities of $n_c$=10$^4$
cm$^{-3}$.

\item \ion{Fe}{2} was assumed to occupy only the ground level because
of the low column density limit observed for the \ion{Fe}{2}* line,
and the high critical densities of $n_c$=10$^6$ cm$^{-3}$ for the
excited levels of the ground state.

\item \ion{Ni}{2} was assumed to occupy only the ground level due to
the high critical density of $n_c$=10$^7$ cm$^{-3}$ for the excited
fine-structure level of the ground state doublet.
\end{enumerate}

The geometrical factor $\zeta$ used in the calculation of the emission
densities is determined in the following manner.  For a rectangular
spectrograph slit that extends across a chord of a spherical nebula,
having its nearest edge at an angular distance $\theta_1$ from the
center and the far edge at $\theta_2$,
\begin{equation}
\label{eqn3}
\zeta = (\pi/3) \, [{\theta_1}^3 - {\theta_2}^3 + 3 {\theta_r}^2 (\theta_2 - \theta_1)]/[\theta_r \, {\theta_o}^2]\,,
\end{equation}
for a nebular shell having angular radius $\theta_r$ with a uniform,
including clumpy, distribution of $n_e$ $n_i$ inside, and assuming the
slit to project across the full extent of the nebula.  In this
circumstance, $\zeta$ is usually only a few tens of percent different
from unity.  By contrast, if the ions are confined within a thin shell
having a thickness much less than $\theta_r$,
\begin{equation}
\label{eqn4}
\zeta = 2 \pi \theta_r (\theta_2 - \theta_1)/{\theta_o}^2\,,
\end{equation}
provided the shell extends beyond $\theta_2$.  Here, $\zeta$ is
independent of the offset from the center of the nebula and can
signify a substantial enhancement of the flux $F_{\lambda}$ over that
which emanates from a similar sized column that is immediately
adjacent to the central star.  Models of IC 418 indicate that its
geometry is that of a thick spherical shell \citep{F69,RW79},
consistent with its fairly uniform surface brightness.  For the slit
position used for our IC 418 observations $\zeta$ should therefore
have values between the extremes represented by eqs.~\ref{eqn3} and
~\ref{eqn4}, resulting in values $1.3<\zeta<3.9$.  We have chosen the
intermediate value $\zeta$=2 for our calculations.  Because of the
uncertainties associated with the column densities due to blending and
line saturation effects, the emission densities derived for most of
the ions span a relatively large range of values.

\subsection{Comparative Abundances and Emission Densities}

Relative abundances of ions can be obtained from either the emission
measures or the column densities by taking simple ratios if the ions
are co-spatial.  Because they are derived from independent techniques,
a comparison of the relative abundances from the different methods
should provide a measure of their reliability.  An alternative way to
test the consistency of the results for an individual ion is to
compare the emission density of that ion with the measured value of
the electron density in the nebular shell.  Any departure of the
$<\!\!n_e\!\!>_{em}$ from the measured electron density indicates
discrepant results between the emission and absorption analyses for
that ion.

\begin{figure}
\figurenum{3} \plotone{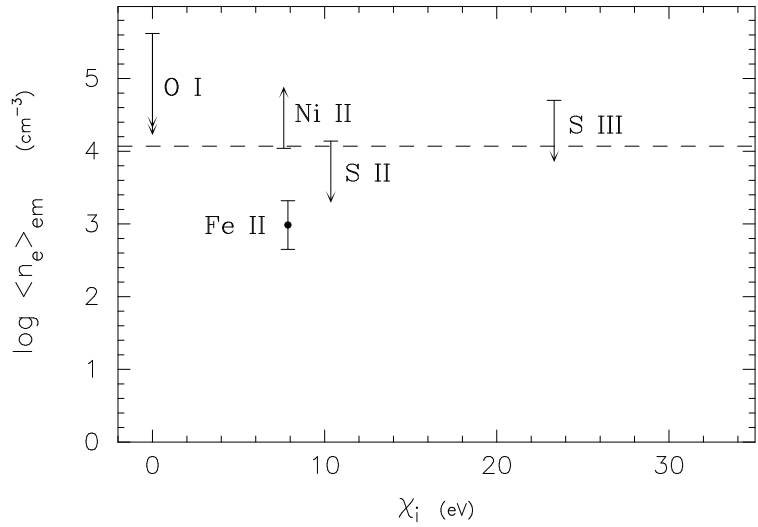} \figcaption[f3g.ps]{The emission
densities of ions in IC 418 plotted in terms of the ionization
potential of the ion.  The dashed line shows the mean electron density
of the nebular gas as determined from the [\ion{O}{2}], [\ion{S}{2}],
and [\ion{Cl}{3}] doublet line ratios. The strongly saturated
\ion{O}{1} absorptions lead to an especially uncertain value of the
emission density for that ion. \label{fig3}}
\end{figure}

Our observations of the doublet ratios of [\ion{O}{2}]
$\lambda$3729/$\lambda$3726, [\ion{S}{2}] $\lambda$6731/$\lambda$6716,
and [\ion{Cl}{3}] $\lambda$5538/$\lambda$5518 can be used to derive
values of $n_e$ of, respectively, $(1.0, 1.0, \textrm{\&} 1.6) \times
10^4$ cm$^{-3}$ for IC 418, in agreement with the results of
\citet{SK89}.  A plot of the emission densities for the five ions for
which we were able to determine values is given in Fig.~\ref{fig3},
shown as a function of the ionization potential required to produce
each ion.  Unfortunately, the quality of the data are such that we
have been able to establish only lower or upper limits to the column
densities, rather than a specified range, and this restricts our
useful data points for IC 418 to a small number.  Clearly, in order to
exploit this technique a much larger database of high resolution UV
spectra of nebular central stars is required than is presently
available, but this situation should soon be rectified with the
current instrumentation on {\sc HST}.

The horizontal dashed line in Fig.~\ref{fig3} represents the mean
electron density of the IC 418 shell as determined from the above
doublet line ratios, and except for \ion{Fe}{2}, which has a notably
low emission density, the allowable ranges of the other emission
densities include the mean value of $n_e \approx 10^4$ for the nebula.
The emission density for \ion{Ni}{2} is a bit high, which might
indicate a possible contribution to the line excitation by a process
other than electron impact, as mentioned previously and elaborated on
below.

Given that the mean value of $n_e$ in IC 418 is close to or above the
upper limits to the emission densities for \ion{S}{2} and \ion{Fe}{2},
the data are suggestive that the emission densities for the low
ionization ions may be generally lower than the electron density.  If
true, it would either require unknown processes which act to
\textit{decrease} the emission line intensities \textit{below} the
collisional excitation rates, or the existence of an appreciable
amount of (presumably cool) gas which absorbs but does not radiate.
The presence of significant amounts of neutral and low ionization gas
associated with ionized nebulae has been documented for various
objects, and it may be a more extensive phenomenon than previously
believed \citep{D95,H96}.

There is an additional feature of Fig.~\ref{fig3} that is important to
understand and which demonstrates the value of the emission density
parameter.  Ions which occupy the same regions of gas should have the
same emission densities.  Stated differently, abundance ratios derived
for co-spatial ions should be invariant to both the line of sight and
whether determined from emission or absorption lines.  The ions Fe$^+$
and Ni$^+$ are expected to be co-spatial because iron and nickel have
very similar ionization potentials for their first five stages of
ionization, and similar condensation temperatures.  In addition, the
ratios of the photoionization cross sections to the recombination
coefficients for the ions of Fe and Ni are very similar over a wide
range of energies except those just above the thresholds (cf. Fig.~3
of Sofia \& Jenkins 1998).  Thus, the gas phase abundance ratios of
Fe$^+$/Ni$^+$ derived from either the emission or the absorption lines
should have similar values.  This is not the case for the IC 418 data.

The column densities derived from the nebular absorption equivalent
widths of \ion{Fe}{2} $\lambda$1260.5 and \ion{Ni}{2} $\lambda$1317.2
yield Fe$^+$/Ni$^+>25$ (Table~\ref{tbl1}), compared to the solar value
of 19.  This ratio is consistent with that found for the ISM, in which
Ni is depleted slightly more than Fe \citep{SS96}.  The relative
emission fluxes of [\ion{Fe}{2}] $\lambda$8617 and [\ion{Ni}{2}]
$\lambda$7378, on the other hand, lead to an abundance ratio of
Fe$^+$/Ni$^+$ = 5, assuming predominant collisional excitation of the
lines (Table~\ref{tbl2}).  The discrepancy between these two values is
greater than can be accounted for by observational uncertainties,
suggesting that either the f-values and collision strengths used in
the calculation are in error or the proper excitation mechanisms have
not been accounted for in the emission calculation.  Collisional
quenching of the lines is not a factor since the critical densities
for their upper levels exceed $10^6$ cm$^{-3}$.

The possibility that the atomic cross sections for the relevant
transitions may be in error by combined factors of five cannot be
ruled out, although this is believed to be unlikely.  A more realistic
possibility is that we have overlooked a significant source of
excitation for the [\ion{Ni}{2}] line.  The complicated structure of
\ion{Fe}{2} makes it likely that our assumption of a 3-level atom
underestimates the excitation of the [\ion{Fe}{2}] line \citep{BP96},
but that only serves to exacerbate the discrepancy by decreasing the
emission-determined Fe$^+$/Ni$^+$ ratio further.
   
The formation of the absorption lines is rather straightforward
because they originate from the ground state, so attention can be
focused on the excitation processes for the emission lines.  There
have been independent suggestions of problems in this regard, as
anomalously high nickel abundances derived for various nebulae from
the [\ion{Ni}{2}] $\lambda$7378 line \citep{H88} have previously led
to the suggestion that this line can be excited predominantly by
fluorescence of stellar continuum radiation.

\citet{LU95} has shown that resonance fluorescence of stellar
continuum radiation can be important in exciting [\ion{Ni}{2}]
$\lambda$7378 near a strong UV source, and \citet{BP96} have extended
that work to a number of [\ion{Ni}{2}] and [\ion{Fe}{2}] lines.  We
have investigated this process for our observations, and at the slit
position we have used resonance scattering does compete with the
collisional excitation of [\ion{Ni}{2}] $\lambda$7378, although it is
not important for the other lines.  Using the observed IUE stellar
continuum flux for IC 418 at 1742\AA\ (the pumping wavelength for the
\ion{Ni}{2} resonance fluorescence) and the parameters derived by
\citet{PCS89} for the central star, we find that at most 25\% of the
observed emission of [\ion{Ni}{2}] $\lambda$7378 should be due to
resonance fluorescence if all of the line emission were to originate
from the slit position corresponding to the minimum distance of
4.3\arcsec\ , roughly $4\times10^{16}$ cm, from the central star.
Since the distribution of the line emission is moderately uniform
along the slit the actual contribution of resonance fluorescence to
the observed $\lambda$7378 flux should be less than 25\%.  Thus,
resonance fluorescence does make a contribution to the [\ion{Ni}{2}]
line, but it fails to account for most of the line flux that we have
observed.

We are left with the following facts regarding the emission and
absorption line strengths in IC 418:

\begin{enumerate}
\item The intensity of [\ion{Fe}{2}] and probably also [\ion{S}{2}]
emission is weaker than it should be for their derived column
densities and the nebular electron density of $n_e=10^4$ cm$^{-3}$.

\item The relative [\ion{Fe}{2}] and [\ion{Ni}{2}] intensities differ
from the values they should have based on the measured Fe$^+$ and
Ni$^+$ column densities, assuming collisional excitation of the lines,
and fluorescent excitation of the [\ion{Ni}{2}] does not account for
this discrepancy.
\end{enumerate}

We have explored various alternatives that might explain the
difference between the Fe$^+$/Ni$^+$ abundance derived from the
emission and the absorption lines, but no satisfactory explanation has
emerged.  The fact that the \ion{Ni}{2} emission density is closer to
the electron density than that of \ion{Fe}{2} suggests that the
[\ion{Fe}{2}] should be considered to be abnormally weak more than the
[\ion{Ni}{2}] is overly strong.  It would be instructive to determine
the emission densities for \ion{Ni}{2} in those nebulae for which
\citet{H88} deduced a high Ni abundance from the [\ion{Ni}{2}]
emission lines in order to resolve this dilemma.

The data from IC 418 show that there are some discrepancies in the
abundances derived for various ions from their absorption versus
emission lines.  Nevertheless, inconsistent line strengths for ions
should not be considered too problematical on the basis of data from
just this one object.  If a pattern emerges for the same ions in other
objects, however, it would signal problems with either the atomic data
or the excitation processes assumed for those ions.  Data from
additional nebulae are necessary to resolve the questions posed here,
and we have performed a search of the electronic edition of the
\citet{A92} catalogue of PNe in order to identify good candidate PNe
for such future studies.  Table~\ref{tbl3} lists those objects which
have relatively bright central stars ($m_v<14.5$), radial velocities
that deviate by more than 40 km s$^{-1}$ from the LSR, and higher
surface brightnesses.  These PNe should be among the prime targets for
followup work.

\section{Summary}

The combination of both absorption and emission analyses provides
important information on the structure of emission regions, physical
conditions in the gas, and element abundances, and should be utilized
in the abundance analyses of nebulae.  It is imperative that UV
spectroscopy of a larger sample of stars embedded in and behind
emission regions be obtained in order to address the reliability of
abundance determinations from absorption versus emission lines.
Because of the different lines of sight and techniques employed, a
combined emission and absorption analysis provides more complete
information about the objects, including their gas distribution and
inhomogeneities, than individual emission or absorption analyses
alone.  The emission density of an ion is an important parameter that
characterizes its emission and absorption properties, and also
provides information on the gas geometry, especially the nature of
inhomogeneities, and line excitation mechanisms.

The combined analyses of IC 418 suggest that some forbidden lines,
particularly those of low ionization, may not lead to correct gas
phase ion abundances using standard assumptions of emission analysis.
This is particularly evident for Fe$^+$.  The data also suggest the
likely existence of non-emitting gas in IC 418 whose presence is
revealed by the absorption lines.  Additional data are needed that
provide more tightly constrained column densities for a larger sample
of objects in order to fully exploit the technique.  However, it is
clear that the comparison of emission- and absorption-line data for
the same object is a important tool for determining abundances from
nebular spectra and for establishing which emission lines are reliable
indicators of ion abundance.

\clearpage

\clearpage

\clearpage

\clearpage

\begin{deluxetable}{
c    % lambda
c    % log flambda
c    % species
c    % W_lambda
c    % log N
}
\tablecolumns{5}
\tablewidth{0pt}
\tablecaption{Ultraviolet Line Results \label{tbl1}}
\tablehead{
\colhead{$\lambda$\tablenotemark{a}} & \colhead{$\log
f\lambda$\tablenotemark{a}} & \colhead{Species} &
\colhead{$W_\lambda\pm 1\sigma$ error} & \colhead{$\log N$}\\
\colhead{(\AA)} & \colhead{} & \colhead{} & \colhead{(m\AA)} &
\colhead{(${\rm cm}^{-2}$)}
}
\startdata
1550.770 & 2.167 & \ion{C}{4} &  50.6$\pm   3.1$& $\gtrsim 13.40$\\
1304.858\tablenotemark{b}& 1.830 & O I*& 115.3$\pm   5.1$&$\gg 14.60$\\
&&& 112.4$\pm   3.2$\\
1306.029 & 1.830 & O I**&  98.9$\pm   2.0$&$\gg 14.53$\\
1239.925 & $-0.106$ & Mg II&  13.6$\pm   1.5$&$15.23-\infty$\\
1240.395 & $-0.355$ & Mg II&  10.1$\pm   1.9$\\
1264.738 & 3.128 & Si II*& 151.7$\pm   2.5$&\tablenotemark{c}\\
1533.431 & 2.303 & Si II*& 130.9$\pm   2.6$\\
1197.394\tablenotemark{b} & 2.242 & Si II*&  93.1$\pm   3.7$\\
&&&  94.9$\pm   5.1$\\
1265.002 & 2.174 & Si II*&  91.7$\pm   3.6$\\
1309.276 & 2.078 & Si II*&  77.0$\pm   1.7$\\
1393.755 & 2.855 & Si IV&  64.7$\pm   1.9$&$13.11-13.28$\\
1402.770 & 2.554 & Si IV&  44.9$\pm   2.2$\\
1253.811 & 1.136 & S II&  17.7$\pm   3.1$&$14.50-\infty$\\
1250.584 & 0.832 & S II&  18.2$\pm   3.4$\\
1194.058 & 1.315 & S III*&  49.4$\pm   2.6$&$14.35-\infty$\\
1200.966\tablenotemark{b}& 1.364 & S III**&  40.8$\pm
2.7$&$14.23-14.82$\\
&&&  42.8$\pm   2.8$\\
1201.726 & 0.616 & S III**&   9.4$\pm   2.0$\\
1260.533 & 1.484 & Fe II&  12.5$\pm   2.2$&$\gtrsim 13.57$\\
1611.200 & 0.339 & Fe II&  $-4.0\pm   4.7$&$< 14.24$\\
1621.686 & 1.823 & Fe II*&   4.5$\pm   4.1$&$<13.12$\\
1317.217\tablenotemark{d} & 2.284 & Ni II&  $-1.8\pm   3.7$&$<12.16$\\
&&&   2.5$\pm   2.3$\\
&&&   2.9$\pm   1.8$\\
1370.132 & 2.254 & Ni II&  $-1.2\pm   1.9$\\
1237.059 & 3.035\tablenotemark{e}& Ge II&   7.2$\pm   2.3$&$\gtrsim
11.78$\\
\\
\\
\enddata
\tablenotetext{a}{Wavelengths and $f$-values from \citet{MO00}, except
for \ion{C}{4} and \ion{Si}{4} which were taken from \citet{MO91}.}
\tablenotetext{b}{This feature is seen twice, at opposite ends of two
adjacent orders of diffraction.}
\tablenotetext{c}{The profile of \ion{Si}{2}* is complex and probably
represents absorptions by many unrelated regions.  For this reason, we
are not supplying a result for the column density.}
\tablenotetext{d}{This spectrum covering the location of this transition
was observed three times: once in one exposure, and at opposite ends of
two adjacent orders of diffraction another exposure.}
\tablenotetext{e}{\citet{MO91} lists this value as 2.035, which is
incorrect.}
\end{deluxetable}

\clearpage

\begin{deluxetable}{lcc@{}r@{.}l@{\extracolsep{1pc}}lr@{.\extracolsep{0pt}}lc}
\tablecolumns{9}
\tablewidth{0pt}
\tablecaption{Ion Emission Densities \label{tbl2}}
\tablehead{
\multicolumn{2}{c}{Line} & \multicolumn{1}{c}{F$_0$} & \multicolumn{2}{c}{$\Omega_{12}$} & \multicolumn{1}{c}{A$_{21}$} & \multicolumn{2}{c}{$EM_i$} & \multicolumn{1}{c}{$<\!\!n_e\!\!>_{em}$} \\
\multicolumn{1}{c}{} & \multicolumn{1}{c}{} & \multicolumn{1}{c}{(erg cm$^{-2}$ sec$^{-1}$)} & \multicolumn{2}{c}{} & \multicolumn{1}{c}{(sec$^{-1}$)} & \multicolumn{2}{c}{(cm$^{-6}$ pc)} & \multicolumn{1}{c}{(cm$^{-3}$)}
}
\startdata
\textrm{[}\ion{O}{1}\textrm{]} & $\lambda 6300$ & $1.03\times 10^{-12}$ & 0&12 & $6.3 \times 10^{-3}$ & 197 & & $\ll 4.2 \times 10^{5}$ \\
\textrm{[}\ion{S}{2}\textrm{]} & $\lambda 6717$ & $9.1\times 10^{-13}$ & 4&8 & $2.6 \times 10^{-4}$ & 1&2 & $< 1.1 \times 10^{4}$ \\
 & $\lambda 6731$ & $1.89 \times 10^{-12}$ & 3&2 & $8.8 \times 10^{-4}$ & 1&7 &  $< 1.7 \times 10^{4}$ \\
\textrm{[}\ion{S}{3}\textrm{]} & $\lambda 9069$ & $8.3 \times 10^{-12}$ & 8&4 & $2.2 \times 10^{-2}$ & 7&1 & $< 4.7 \times 10^{4}$ \\
\textrm{[}\ion{Fe}{2}\textrm{]} & $\lambda 8617$ & $4.5 \times 10^{-15}$ & 0&93 & $2.7 \times 10^{-2}$ & 0&025 & $(0.45-2.1) \times 10^{3}$ \\
\textrm{[}\ion{Ni}{2}\textrm{]} & $\lambda 7378$ & $1.93 \times 10^{-15}$ & 0&75 & $2.3 \times 10^{-1}$ & 0&0052 & $> 1.1 \times 10^{4}$
\enddata
\end{deluxetable}

\clearpage

\begin{deluxetable}{llcrrrrc}
\tablecolumns{8}
\tablewidth{0pt}
\tabletypesize{\scriptsize}
\tablecaption{Candidate PNe for Absorption+Emission Studies \label{tbl3}}
\tablehead{
\colhead {PN G} & \colhead{Common Name} & \multicolumn{1}{c}{RA (J2000)} & \multicolumn{1}{c}{Dec (J2000)} & \multicolumn{1}{c}{Vel(LSR)} & \multicolumn{1}{c}{$V_*$} & \multicolumn{1}{r}{Diam (\arcsec)} & \colhead{SB (H$\beta$)\tablenotemark{a}}
}
\startdata
118.0-08.6 & Vy 1-1 & 00:16:00 & +53:36:00 & -42.9 & 14.2 & 5.2 & 1.4$\times 10^{-13}$ \\
118.8-74.7 & NGC 246 & 00:44:32 & -12:08:44 & -51.1 & 12.0 & 245.0 & 6.3$\times 10^{-16}$ \\
165.5-15.2 & NGC 1514 & 04:06:08 & +30:38:43 & 51.9 & 9.4 & 132.0 & 7.7$\times 10^{-16}$ \\
215.2-24.2 & IC 418 & 05:25:09 & -12:44:15 & 42.1 & 10.2 & 12.0 & 2.3$\times 10^{-12}$ \\
197.8+17.3 & NGC 2392 & 07:26:13 & +21:00:51 & 63.4 & 10.5 & 19.5 & 1.4$\times 10^{-13}$ \\
320.1-09.6 & He 2-138 & 15:51:19 & -66:00:26 & -49.7 & 10.9 & 7.0 & 5.2$\times 10^{-13}$ \\
341.6+13.7 & NGC 6026 & 15:58:07 & -34:24:16 & -96.5 & 13.3 & 40.0 & 1.8$\times 10^{-15}$ \\
326.0-06.5 & He 2-151 & 16:11:25 & -59:46:34 & -129.2 & 13.1 & 3.0 & 1.6$\times 10^{-13}$ \\
325.8-12.8 & He 2-182 & 16:49:49 & -64:09:39 & -92.8 & 13.4 & 3.0 & 1.6$\times 10^{-12}$ \\
353.7+06.3 & M 2-7 & 17:02:02 & -30:28:24 & -46.7 & 14.0 & 7.8 & 5.3$\times 10^{-15}$ \\
345.0-04.9 & Cn 1-3 & 17:22:34 & -44:08:54 & -72.7 & 14.3 & 5.0 & 4.1$\times 10^{-13}$ \\
345.2-08.8 & Tc 1 & 17:41:52 & -46:04:10 & -78.2 & 11.4 & 9.6 & 3.1$\times 10^{-13}$ \\
096.4+29.9 & NGC 6543 & 17:58:34 & +66:38:05 & -50.1 & 11.1 & 19.5 & 8.2$\times 10^{-13}$ \\
002.4-03.7 & M 1-38 & 18:02:55 & -28:40:54 & -59.6 & 14.4 & 3.3 & 1.2$\times 10^{-13}$ \\
342.5-14.3 & Sp 3 & 18:03:22 & -51:01:35 & 48.6 & 12.6 & 35.5 & 8.0$\times 10^{-15}$ \\
003.3-04.6 & Ap 1-12 & 18:08:25 & -28:23:21 & 162.5 & 13.3 & 12.0 & 3.5$\times 10^{-14}$ \\
011.7-00.6 & NGC 6567 & 18:10:48 & -19:05:13 & 132.4 & 14.4 & 7.6 & 2.6$\times 10^{-13}$ \\
002.0-06.2 & M 2-33 & 18:11:53 & -30:16:32 & -102.1 & 14.4 & 5.8 & 9.5$\times 10^{-14}$ \\
016.4-01.9 & M 1-46 & 18:25:04 & -15:34:53 & 43.9 & 12.8 &  11.0 & 3.3$\times 10^{-14}$ \\
015.4-04.5 & M 1-53 & 18:32:53 & -17:38:38 & 76.3 & 14.3 & 6.0 & 2.2$\times 10^{-14}$ \\
003.9-14.9 & Hb 7 & 18:52:23 & -32:19:49 & -56.4 & 14.0 &  4.0 & 4.5$\times 10^{-13}$ \\
082.5+11.3 & NGC 6833 & 19:48:20 & +48:50:01 & -91.0 & 14.5 & 2.0 & 1.8$\times 10^{-12}$ \\
054.1-12.1 & NGC 6891 & 20:12:48 & +12:32:54 & 58.9 & 12.4 &  15.0 & 1.3$\times 10^{-13}$ \\
058.3-10.9 & IC 4997 & 20:17:51 & +16:34:22 & -49.5 & 14.4 & 1.6 & 1.5$\times 10^{-11}$ \\
093.4+05.4 & NGC 7008 & 20:59:05 & +54:20:41 & -58.8 & 13.2 & 86.0 & 2.4$\times 10^{-15}$
\enddata
\tablenotetext{a}{Surface brightness in units of absolute H$\beta$ flux per square arcsec}
\end{deluxetable}

\end{document}